\documentstyle[sprocl]{article}

\newcommand{\be}{\begin{equation}}
\newcommand{\ee}{\end{equation}}
\begin{document}
\title{LIGHT-LIGHT AND HEAVY-LIGHT MESONS SPECTRA FROM NONPERTURBATIVE QCD}
\author{Alexei Nefediev\\
{\em Institute of Theoretical and Experimental Physics, Moscow
117218, Russia and\\
Centro de F\'\i sica das Interac\c c\~oes Fundamentais (CFIF)
Departamento de F\'\i sica, Instituto Superior T\'ecnico
Av. Rovisco Pais, P-1049-001 Lisboa, Portugal}}
\maketitle
\begin{abstract}
Properties of light-light mesons are described by the effective
Hamiltonian with spinless quarks derived from QCD. The spectrum is
computed by the WKB method and shown to reproduce the celebrated
linear Regge trajectories even for the lowest levels. The correct
string slope of the trajectories naturally appears in the present
approach as the string dynamics is taken into account properly.

Similar method is applied to heavy-light mesons and a set of
corrections to the Hamiltonian is taken into account including
spin-spin and Tomas spin-orbit interactions. The numerical results
for the spectrum are compared with the experimental data and with
the results of recent lattice calculations.
\end{abstract}
\section{Introduction}
One of the most successful models of confinement in QCD is the string picture which
exploits the idea of the flux tube formation between the colour 
constituents in hadrons. 
The small radius of the string compared to the hadronic size makes it possible to
construct quantum mechanical quark models with the interquark interaction
described by either non-relativistic \cite{isgur} or relativistic string
(see {\it e.g.} \cite{DKS}). 
The role of the string becomes
especially important if light quarks are involved, so that the proper
string dynamics should be taken into account together with the quarks one
when studying the properties of hadrons.

\section{Light-light mesons}

Starting from the gauge invariant Green's function of the $q\bar q$ system,
neglecting spins and
using the Feynman-Schwinger representation for the one-particle propagators as well
as the area law asymptotic for the Wilson loop one arrives at the following
Lagrangian of the system
\be
L(\tau)=-m_1\sqrt{\dot{x}_1^2}-m_2\sqrt{\dot{x}_2^2}
-\sigma\int_0^1d\beta\sqrt{(\dot{w}w')^2-\dot{w}^2w'^2},
\label{1}
\ee
where the interaction is described by the Numbu-Goto term for the minimal 
string bounded by the quarks trajectories \cite{DKS}. 
We use the straight-line anzatz for the minimal string 
$w_{\mu}(\tau,\beta)=\beta x_{1\mu}+(1-\beta)x_{2\mu}$ \cite{DKS}. 
Introducing einbein fields $\mu$ (dynamical mass of the quark) and $\nu$
(density of the string energy) to get rid of the square roots in (\ref{1}) and
using the standard techniques, one finds the Hamiltonian of
the system in the form (we consider equal quark masses) \cite{DKS}
\be
H=\frac{p^2_r+m^2}{\mu(\tau)}+
\mu(\tau)+\frac{\hat L^2/r^2}
{\mu+2\int^1_0(\beta-\frac{1}{2})^2\nu(\beta) d\beta}+
\int^1_0\frac{d\beta}{2}\left(\frac{\sigma^2r^2}{\nu(\beta)}+\nu(\beta)\right).
\label{3}
\ee

Getting rid of the einbein $\nu$ by taking extremum in Hamiltonian
(\ref{3}) and keeping  the other einbein $\mu$ as a variational parameter
$\mu_0$ one finds
\be
H=\frac{p^2_r+m^2}{\mu_0}+\mu_0+U(\mu_0,r),
\label{5}
\ee 
where the effective potential $U$ has a rather complicated form; its 
dependence on $\mu_0$ reflects the
nonlocal string-type character of the interaction introduced in (\ref{1}).
Nonrelativistic expansion of (\ref{5}) gives for the interquark potential
\be
V(r)=U(m,r)=\sigma r-\frac{\sigma \hat{L}^2}{6m^2r}+\ldots,
\label{4}
\ee
where the correction to the confining potential is known as
the string one \cite{string_correction,DKS}. 

The spectrum of the Hamiltonian (\ref{5}) is found by the
quasiclassical method and each eigenenergy is minimized tuning
$\mu_0$. Numerical results \cite{MNS} reproduce straight-line Regge
trajectories in the angular momentum $l$ with the inverse slope 
$2\pi\sigma$ \cite{MNS}. 
The difference between $2\pi\sigma$ and the overestimated value
$8\sigma$ found for the Bethe-Salpeter equation with linear confinement is 
entirely due to the proper string dynamics missing in the latter case.
\section{Heavy-light mesons}
Now we apply similar approach to the heavy-light mesons spectrum. The 
zeroth approximation for the Hamiltonian 
with the Coulombic potential and the constant term added and the string
correction to it read $(\kappa=\frac43\alpha_s)$:
\be
H_0=\sum_{i=1}^2\left(\frac{\vec{p}^2+m_i^2}{2\mu_i}+\frac{\mu_i}{2}\right)+\sigma
r-\frac{\kappa}{r}-C_0, V_{str}=-\frac{\sigma
(\mu_1^2+\mu_2^2-\mu_1\mu_2)}{6\mu_1^2\mu_2^2}\frac{\vec{L}^2}{r},
\label{H0}
\ee
whereas other corrections are spin-dependent and coincide in form with the 
Eichten--Feinberg--Gromes results \cite{EFG} up to the change $m_i\to\mu_i$ 
in the denominators \cite{Vsd} (note that for a light quark
$\mu_i\sim 500\div 800MeV\gg m_i$):
$$
V_{sd}=\frac{8\pi\kappa}{3\mu_1\mu_2}(\vec{S}_1\vec{S}_2)
\left|\psi(0)\right|^2-\frac{\sigma}{2r}\left(\frac{\vec{S}_1\vec{L}}{\mu_1^2}+
\frac{\vec{S}_2\vec{L}}{\mu_2^2}\right)
+\frac{\kappa}{r^3}\left(\frac{1}{2\mu_1}+\frac{1}{\mu_2}\right)
\frac{\vec{S}_1\vec{L}}{\mu_1}
$$
$$
+\frac{\kappa}{r^3}\left(\frac{1}{2\mu_2}+\frac{1}{\mu_1}\right)
\frac{\vec{S}_2\vec{L}}{\mu_2}
+\frac{\kappa}{\mu_1\mu_2r^3}\left(3(\vec{S}_1\vec{n})
(\vec{S}_2\vec{n})-(\vec{S}_1\vec{S}_2)\right)+\frac{\kappa^2(\vec{S}\vec{L})}{2\pi\mu^2r^3}
$$
\be
+(2-{\rm ln}(\mu r)-\gamma_E),\quad \gamma_E=0.57.\hspace*{5cm}
\label{11}
\ee

Numerical results for the $D$, $D_s$, $B$ and $B_s$ mesons spectra calculated in the 
the given technique with standard parameters \cite{yua,we} 
are in a good agreement with the experimental and lattice data.

In conclusion let us briefly discuss the situation with the ${D^*}'$ resonance
recently claimed by DELPHI Collaboration \cite{DELPHI}. It was reported to have
the mass $2637\pm 6MeV$ that agrees with the predictions of the quark models for
the first radial excitation $2^3S_1(0^-)$ of the $q\bar q$ pair (our prediction
for this
state is $2664MeV$) but its surprisingly small width of about $15MeV$ is in a
strong contradiction with the theoretical estimates \cite{width}. 
Meanwhile it was observed \cite{width} that orbitally
excited states $2^-$ and $3^-$ could have such a small width. Our model
predictions for these states are $2663MeV$ and $2654MeV$ correspondingly, {\it i.e.} they lie 
even lower than the radially excited one. This resolves the main objection to
the identification of the ${D^*}'$ with orbital excitations. Indeed, in quark
models $2^-$ and $3^-$ states lie at least $50-60MeV$ higher that the experimentally 
observed value. In our approach extra negative contribution 
to the masses of orbitally excited states is readily delivered by the string
correction discussed above, which comes from the proper string dynamics inside
meson.

Financial support of INTAS-RFFI grant IR-97-232 and RFFI grants 00-02-17836 and 00-15-96786
is gratefully acknowledged.

\end{document}